\documentclass[prl,twocolumn,aps,showpacs]{revtex4}
\usepackage{graphicx}
\usepackage{amsmath}
\usepackage{subfigure}
\begin{document}

\title{Electronic excitation spectrum of metallic carbon nanotubes}

\author{S. Sapmaz, P. Jarillo-Herrero, J. Kong, C. Dekker, L. P. Kouwenhoven, and H. S. J. van der Zant}
\affiliation{ Kavli Institute of Nanoscience, Delft University of
Technology, PO Box 5046, 2600 GA Delft, The Netherlands}

\begin{abstract}
We have studied the discrete electronic spectrum of closed
metallic nanotube quantum dots. At low temperatures, the stability
diagrams show a very regular four-fold pattern that allows for the
determination of the electron addition and excitation energies.
The measured nanotube spectra are in excellent agreement with
theoretical predictions based on the nanotube band structure. Our
results permit the complete identification of the electron quantum
states in nanotube quantum dots.
\end{abstract}

\maketitle

Since their discovery~\cite{Iijima} carbon nanotubes (NTs) have
emerged as prototypical one-dimensional conductors~\cite{review}.
At low temperatures, NT devices form quantum dots (QDs) where
single-electron charging and level quantization effects
dominate~\cite{Tans,Bockrath}. A continuous improvement in device
fabrication and NT quality has enabled the recent observation of
two-electron periodicity in 'closed' QDs~\cite{Cobden} and
four-electron periodicity in 'open' single- and multi-wall NT
QDs~\cite{Liang,Buitelaar}. Theoretically, the low-energy spectrum
of single wall nanotube (SWNT) QDs has been modeled by Oreg
\emph{et al.},~\cite{Oreg}. Experiments on open NT QDs are
compatible with this model, but the presence of the Kondo effect
and broadening of the energy levels prevents the observation of
the full spectrum~\cite{NoteKondo}. An analysis of the electronic
excitations is therefore still lacking.

\begin{figure}[ht!]
\includegraphics[angle=0,width=0.73\linewidth]{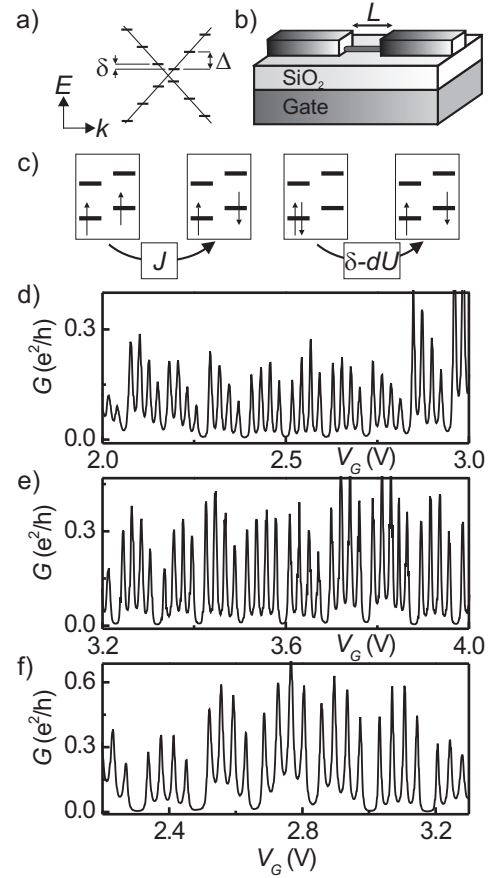}
\caption{\label{CBtraces} (a) Low-energy band structure of a
metallic SWNT. In a finite length SWNT, the wave vector $k$ is
quantized along the tube axis which leads to a set of quantized
energy levels separated by $\Delta$ in each band. $\delta$ is the
mismatch between the two bands. (b) Schematic diagram of the
device geometry. (c) Meaning of $J$ (left) and $dU$ (right). The
exchange energy favors spin alignment and $dU$ is the extra
charging energy associated with placing two electrons in the same
energy level. (d), (e), and (f) Conductance as a function of gate
voltage in the linear response regime at 4 K for three different
CVD grown samples. The NT lengths are 500, 680 and 760 nm,
respectively.}
\end{figure}

\begin{figure*}[ht]
\includegraphics[angle=0,width=0.95\textwidth]{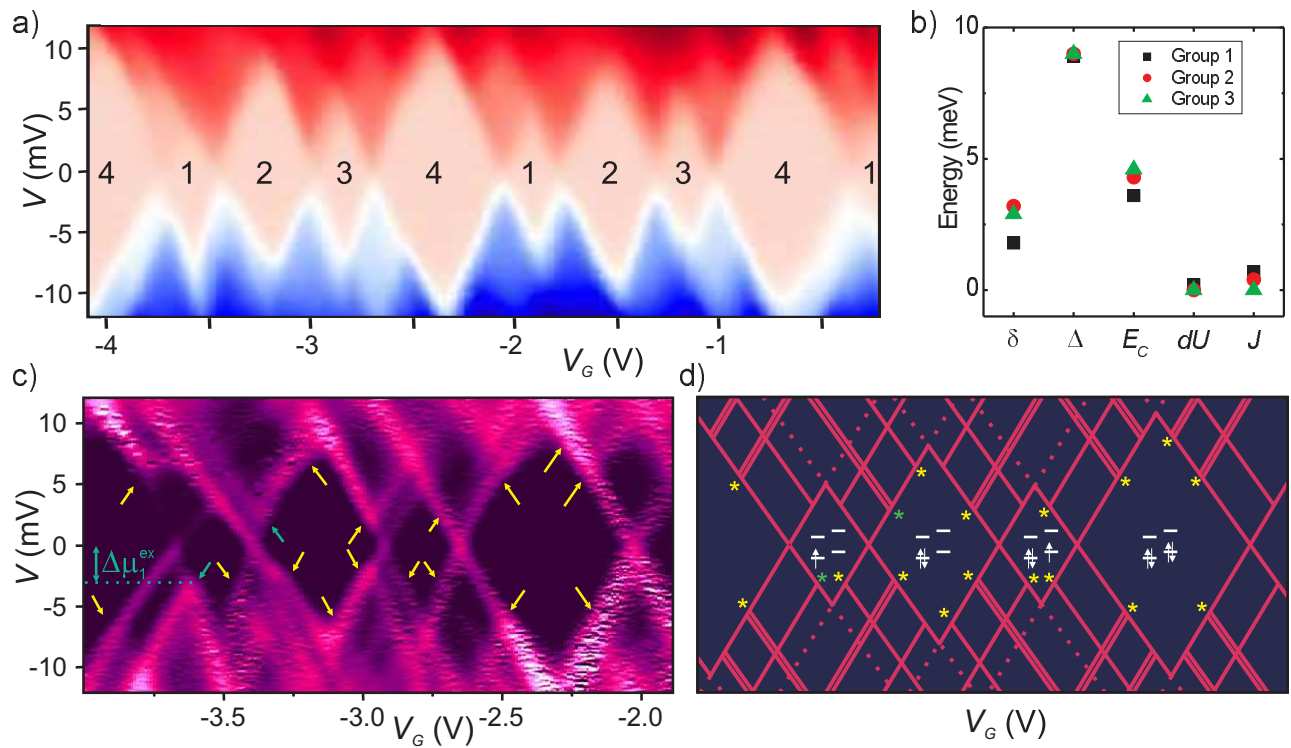}
\caption{\label{SampleA} Sample A - Color plot of the current as a
function of $V$ and $V_G$ at $T = 300$~mK. Blue corresponds to
$-40$~nA and red to $+40$~nA.(b) Values of the parameters for
three different groups of four (see text).(c) The differential
conductance ($dI/dV$) for the first group from (a). Black is zero
and bright is $>$12 $\mu$S. Lines running parallel to the diamond
edges correspond to discrete energy excitations. The excitation
energies corresponding to the green arrows have been used to
deduce the model parameters. The predicted excitations are
indicated by yellow. (d) Calculated spectrum for sample A. The
stars correspond to the arrows in (c). The white diagrams indicate
the spin filling of the ground state.}
\end{figure*}

\begin{figure*}[ht]
\includegraphics[angle=0,width=0.95\textwidth]{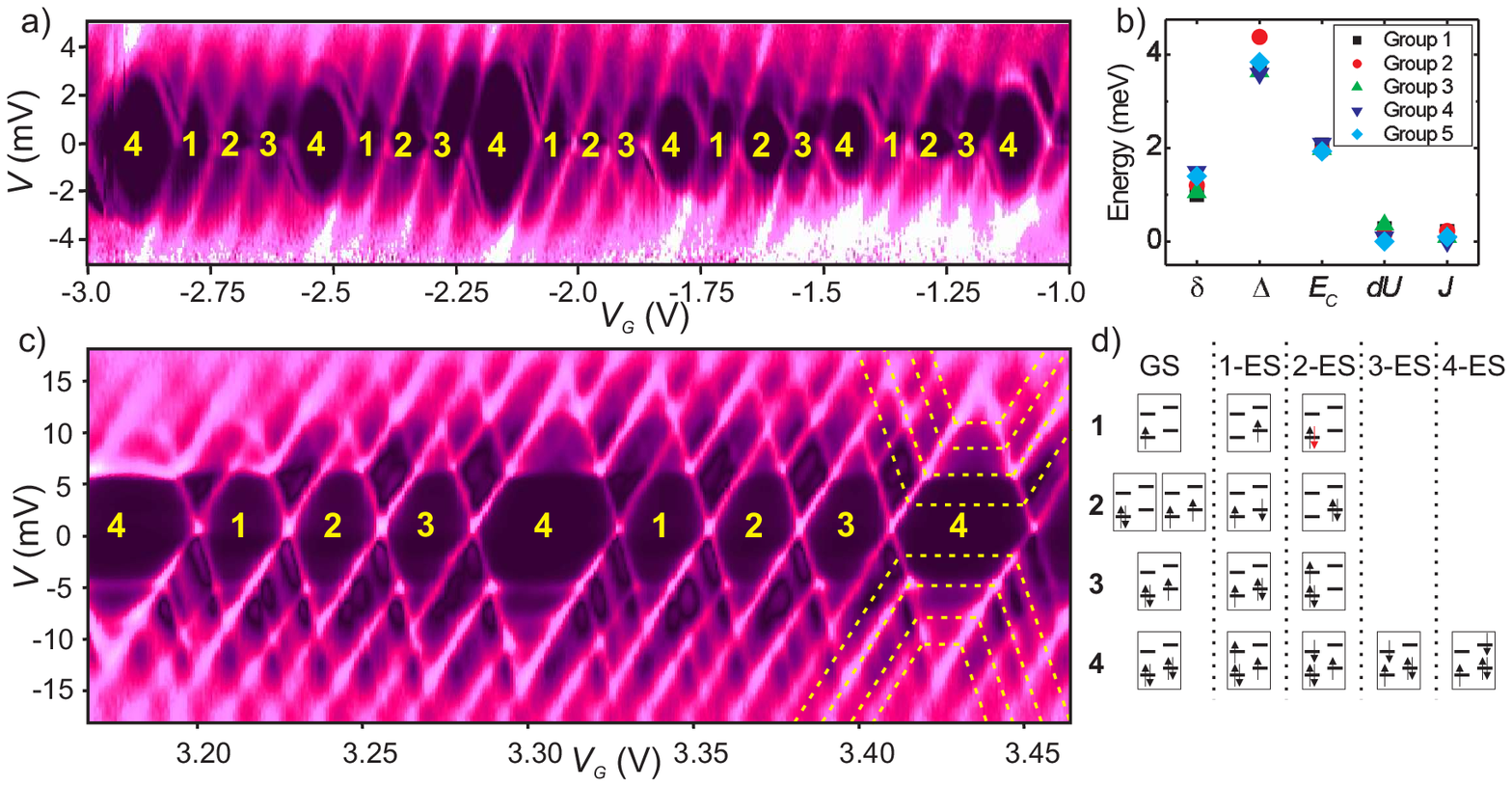}
\caption{\label{SampleB} Differential Conductance of sample B (a)
and C (c) as a function of $V$ and $V_G$ measured at 300~mK. Black
represents $dI/dV \sim 0$, while lighter tones correspond to a
higher conductance. Yellow lines in (c) indicate the excited
states together with inelastic cotunneling. (b) Obtained
parameters for sample B. (d) Electron quantum states of the NT QD.
The numbers on the left denote the ground state (GS) number of
electrons in the last occupied shell. The left column indicates
the GS electron configuration (note that the two-electron GS is
degenerate). Columns on the right denote the excited state (ES)
configuration. Up to four ES are visible in the large Coulomb
diamonds~\cite{NoteDEG}. The red arrow in the second ES for one
electron corresponds to an electron excited from the lower shell.}
\end{figure*}

The two-fold degenerate, low-energy band structure of a metallic
SWNT is schematically shown in Fig.~\ref{CBtraces}a. Quantization
along the nanotube axis leads to a set of single particle states
that are equally spaced because of the linear dispersion
relation~\cite{Dresselhaus Book}. The combination of the two bands
and the spin yields a four-fold periodicity in the electron
addition energy. The simplest model to describe QDs is the
Constant Interaction (CI) model~\cite{LeoReport}, which assumes
that the charging energy is constant and independent of the
occupied single particle states. To describe NT QDs the CI-model
has been extended~\cite{Oreg} to include five independent
parameters: the charging energy $E_C$, the quantum energy level
separation $\Delta$, the subband mismatch $\delta$ (see
Fig.~\ref{CBtraces}a), the exchange energy $J$ and the excess
Coulomb energy $dU$. Fig.~\ref{CBtraces}c illustrates the meaning
of the last two parameters. An independent verification of the
Oreg model~\cite{Oreg} requires the observation of the ground
state addition energies and of at least two excited states. Such a
study has not been reported.

In this letter we investigate the excitation spectrum of closed
SWNT QDs. Not only the ground but also the complete excited state
spectrum of these QDs has been measured by transport-spectroscopy
experiments, enabling us to determine all five parameters
independently. With these, the remaining measured excitation
energies are well predicted leading to a complete understanding of
the spectrum, without adjustable parameters.

HiPco~\cite{HiPCO} and CVD~\cite{CVD} grown NTs were used for the
fabrication of the devices. HiPco tubes were dispersed from a
dichloroethane solution on an oxidized, $p$-doped Si substrate.
The CVD nanotubes were grown from catalyst particles on predefined
positions. Individual NTs were located by atomic force microscopy
(AFM) with respect to predefined marker positions. Electrodes are
designed on top of straight segments of NTs. The highly doped
silicon is used as a backgate to change the electrostatic
potential of the NT QD (see Fig.~\ref{CBtraces}b). We have
fabricated NT devices with lengths in between contacts, $L$,
varying from 100 nm to 1 $\mu$m.

Four-electron shell filling is observed in over 15 samples. In
some cases the four-fold pattern extended over more than 60
electrons added to the QD. Figs.~\ref{CBtraces}d-f show
representative examples of Coulomb Blockade (CB)
oscillations~\cite{Grabert} in the linear response regime.
Clearly, the Coulomb peaks are grouped in sets of four reflecting
the two-fold character of the NT bandstructure.

In the following, we focus on three different devices exhibiting
similar four-fold periodicity in CB oscillations. These samples
(A, B and C) had high enough contact resistances so that not only
the electron ground states but also their excited states could be
resolved. Together they provide enough information to determine
all the parameters in the model. We discuss the results of these
three samples separately.

\emph{Sample A}- This device is made from a HiPco NT~\cite{HiPCO}
with $L=180$~nm and a diameter of 1.1~nm as determined by AFM. It
is contacted by evaporating Cr/Au~(5/75~nm) electrodes.
Fig.~\ref{SampleA}a shows the current, $I$, as a function of
source-drain bias voltage, $V$, and gate voltage, $V_G$. In the
light-colored diamond-shaped regions, the current is blocked due
to CB and the number of electrons is fixed. The clear four-fold
periodicity makes it possible to assign the number of electrons in
the last occupied shell. The sizes of the diamonds form an
interesting pattern, namely a repetition of
small/medium/small/big. This pattern is a consequence of the large
subband mismatch compared to the exchange energy, as we show
below.

The addition energy is defined as the change in electrochemical
potential ($\Delta \mu_N$) when adding the ($N+1$) charge to a
quantum dot already containing $N$ charges~\cite{LeoReport}. The
addition energy is obtained by multiplying the diamond width,
$\Delta V_G$, by a conversion factor, $\alpha$ $(\approx 0.017)$,
which relates the gate voltage scale to the electrochemical
potential~\cite{Grabert}.

The Oreg-model yields the following equations for the addition
energy of the $N$-th electron added~\cite{assumption}:
\begin{align}
    \Delta\mu_1=\Delta\mu_3=& E_C+dU+J \label{mu1} \\
    \Delta\mu_2=& E_C+\delta-dU \label{mu2} \\
    \Delta\mu_4=& E_C+\Delta-\delta-dU. \label{mu4}
\end{align}
To extract all five parameters, two more equations are needed.
These are provided by the excitation spectrum. In
Fig.~\ref{SampleA}c we show the numerical derivative of
Fig.~\ref{SampleA}a (i.e., the differential conductance) for the
first group of four. Excited states of the electrons are visible
for all diamonds. The value of a particular excitation energy
equals the bias voltage at the intersection between the excitation
line and the Coulomb diamond edge (see Fig.~\ref{SampleA}c). The
green arrows in diamond one and two in Fig.~\ref{SampleA}c
correspond to the first excitation for one and two electrons extra
on the NT QD respectively. The theoretical values of these two
energies are
\begin{align}
    \Delta\mu_1^{\text{ex}}=& \delta \label{mu1ex} \\
    \Delta\mu_2^{\text{ex}}=& \delta-J-dU. \label{mu2ex}
\end{align}

Equations~(\ref{mu1})-(\ref{mu2ex}) allow us to uniquely determine
the five unknown parameters from the experimental data alone. We
find $E_C=4.3$~meV, $\Delta=9.0$~meV, $\delta=3.2$~meV,
$J=0.4$~meV and $dU \approx 0$~meV. The values of the parameters
do not vary significantly between the different groups, as shown
in Fig.~\ref{SampleA}b. The theoretically expected value for the
level spacing is $\Delta=hv_F/2L$~\cite{Tans}. With $v_F=8.1\cdot
10^5$ m/s~\cite{Serge} and $L=180$~nm, we find 9.3~meV in
excellent agreement with the experimental value.

Figure~\ref{SampleA}d shows the calculated spectrum of the NT QD
using the parameters deduced from the experiment. Some excitations
are split by the exchange energy. The stars in the calculated
spectrum correspond to the arrows in the experimental data. Green
indicates the excitations used in the calculation whereas yellow
denotes the predicted excited states. The calculated spectrum
resembles the measured one strikingly well.

\emph{Sample B}- This sample is CVD grown~\cite{CVD} with a
diameter of 1.3 nm and $L=500$ nm defined by Cr/Au contacts~(5/40
nm). After contacting, the entire NT segment in between electrodes
is suspended by etching away part of the SiO$_2$~\cite{Nygard}. We
have measured the differential conductance, $dI/dV$, as a function
of $V$ and $V_G$ at 300 mK (Fig.~\ref{SampleB}a). Again regular
four-fold patterns are visible in the Coulomb diamonds.

The evolution of the Coulomb peaks as a function of the magnetic
field (not shown here) gives information about the spin filling of
the states~\cite{Josh}. We find that the filling is the same as
sample A. Excited states of the QD are visible in all groups of
four. The model parameters have been extracted using the same
analysis as described above. The result is shown in
Fig.~\ref{SampleB}b. The average values are $E_C=2.0$~meV,
$\delta=1.2$~meV, $J=0.1$~meV, $dU=0.2$~meV and $\Delta=3.0$~meV.
The value of $\Delta$ corresponds to a length of
440~nm~\cite{Tans}, in good agreement with the NT length between
contacts. Furthermore, in all groups of four at least one more
excitation remains for a comparison between theory and experiment.
In all cases we find good agreement~\cite{weak excitation}.

\emph{Sample C}- This NT is CVD grown \cite{CVD} with a diameter
of 2.7~nm and $L=750$~nm. The contacts are made by evaporating
Ti/Au~(20/40 nm). Fig.~\ref{SampleB}c shows $dI/dV$ as a function
of $V$ and $V_G$. A very regular pattern of Coulomb diamonds with
four-fold periodicity is displayed together with the excited
states. In addition, up to three inelastic co-tunneling
lines~\cite{Silvano} are visible (horizontal lines inside the
Coulomb diamonds in Fig.~\ref{SampleB}c).

The observation of three equally sized small diamonds and the fact
that the excitations have the same energy for all four charge
states indicate that $\delta \approx J + 2dU$. We find
$E_C=6.6$~meV, $\Delta=8.7$~meV, $\delta \approx J =2.9$~meV, and
$dU \approx 0$~meV. Theoretically a level separation of 8.7~meV
indicates a NT QD length of $\sim$~200~nm, while the distance
between contacts is 750~nm. This may suggest that sample C
consists of a QD with NT leads connecting it to the contacts. This
is consistent with the large value for $E_C$. Remarkably, all the
predicted excitation lines are present in the
spectrum~\cite{NoteMu4}. Therefore all the electron states can be
assigned (Fig.~\ref{SampleB}d).

In summary, we have presented a complete analysis of the
electronic spectrum in closed NT QDs. Samples with different
lengths, production process (CVD and HiPco) and contact material
all exhibit four-fold periodicity in the electron addition energy.
The very regular Coulomb traces and stability diagrams enable the
determination of the ground and excited state electron energies.
Knowing precisely the spectrum of nanotube quantum dots is of
fundamental importance in experiments involving the application of
high frequency radiation such as photon-assisted tunneling and
coherent control of the electron quantum states.

We thank R. E. Smalley and coworkers for providing the
high-quality HiPco nanotubes, and C. Meyer, W. Wetzels, M.
Grifoni, R. Hanson, K.A. Williams, Yu. V. Nazarov and S. De
Franceschi for discussions. Financial support is obtained from the
Dutch organization for Fundamental Research on Matter (FOM), which
is financially supported by the 'Nederlandse Organisatie voor
Wetenschappelijk Onderzoek' (NWO).


\begin{thebibliography}{4-fold}
\bibitem{assumption} This set of equations corresponds to singlet ground state for
$N=2$. The triplet case is incompatible with experimental data.
\bibitem{NoteMu4} The experimental value of the addition energy for the large Coulomb
diamond is $\sim$15\% larger than deduced from the spectrum. The
small Coulomb diamonds and the complete excitation spectrum is in
perfect agreement with theory.
\bibitem{NoteDEG} There is more than one degenerate configuration for some of the excited
states. For simplicity we only show one of them.
\end{thebibliography}
\end{document}